\documentclass[12pt]{amsart}
\usepackage{array,longtable}
\usepackage[utf8]{inputenc}
\usepackage{bbm}
\usepackage{amssymb,latexsym,fancyhdr,color,graphics}
\usepackage{amsmath, amsthm}
\usepackage{setspace}
\usepackage[dvips]{graphicx}
\usepackage[pdftex]{hyperref}
\usepackage[parfill]{parskip}
\newtheorem{lem}{\bf Lemma}
\newtheorem{asm}{\bf Assumption}

\newtheorem{prop}{\bf Proposition}
\newtheorem{thm}{\bf Theorem}

\newtheorem{rem}{\bf Remark}

\newcommand{\pf}{\mbox{\bf Proof:}\hspace{3mm}}

\definecolor{titlepagegray}{gray}{0.8}

\begin{document}

\title[Utility maximisation]{Expected Utility Maximisation for Exponential Levy Models with option and  information processes}
\maketitle

\begin{center}

{\large  Lioudmila Vostrikova}
\footnote{$^1${This work is supported in part by  ANR-09-BLAN-0084-01 of the Department of Mathematics of Angers's University.}}
\\\vspace{0.2in}
LAREMA,  D\'epartement de
Math\'ematiques, Universit\'e d'Angers, \\2 Bd. Lavoisier - 49045,
{\sc Angers Cedex 01.}
\end{center} 
\vspace{0.1in}
\begin{abstract}We consider expected utility maximisation problem for exponential Levy models  and HARA utilities in presence of illiquid asset in portfolio . This illiquid asset is modelled by an option of European type on another risky asset which is correlated with the first one. Under some hypothesis on Levy processes, we give the expressions of information processes figured in maximum utility formula. As applications, we consider  Black-Scholes models with correlated Brownian Motions, and  also  Black-Scholes models with jump part  represented by Poisson process.  
\end{abstract}

%%%%%%%%%%%%%%%%%%%%%%%%%%%%%%%%%%%%%%%%%%%%%%%%%%%%%%%%%%%%%%%%%%%%%%%%%%%%%%
\begin{section}{Introduction}\label{s1}%%%%%%%%%%%%%%%%%%%%%%%%%%%%%%%%%%%%%%%%
%%%%%%%%%%%%%%%%%%%%%%%%%%%%%%%%%%%%%%%%%%%%%%%%%%%%%%%%%%%%%%%%%%%%%%%%%%%%%%

\par Levy processes was used in Mathematical Finance since a long time. These models
contain a number of popular jump models including General Hyperbolic models and Variance-Gamma models. The use of such processes for modelling allows an excellent fit both for daily log return and intra-day data. The class of Levy processes is also flexible enough to allow the processes with finite and infinite variation, and also with finite and infinite activity. Levy models are not only excellent to fit the data but also mathematically tractable (see \cite{E}, \cite{EK} and references there).
\par Let $X=(X_t)_{t\geq 0}$ be a d-dimensional Levy process, $d\geq 1$, with generating triplet $(b,c,K)$ where $b\in \mathbb{R}^d$ is drift parameter, $c$ is $d\times d$ matrix  related with  continuous  martingale part of $X$ and $K$ is Levy measure which satisfies:
\begin{equation}\label{0}
\int _{\mathbb{R}^d} (|\!|x|\!|^2\wedge 1)K(dx) < \infty .
\end{equation}
 \par As known, the law of such process is entirely determined by its one-dimensional distributions and  the characteristic function of $$X_t= \,^{\top}(X^1_t, X^2_t,\cdots, X^d_t)$$ at $\lambda\in \mathbb{R}^d$ is given by:
$$\phi _{X_t}(\lambda)= \exp \{t\psi (\lambda)\}$$
where the characteristic exponent of $X$
$$\psi (\lambda)= \exp\left\{i\,^{\top}\lambda  b - \frac{1}{2}\,^{\top}\lambda c \lambda +
\int _{\mathbb{R}^d} (e^{i^{\top}\!\lambda x} -1- i\,^{\top}\!\lambda \,l(x) ) K (dx)\right\}$$
 with the truncation function $l$.
In general,  truncation function $l:\mathbb{R}^d\rightarrow \mathbb{R}^d$ is a bounded function with compact support such that $l(x)=x$ in the neighbourhood of zero. The classical choice of $l$ is $l(x) = x {\bf 1}_{\{|\!|x|\!|\leq 1\}}$ where ${\bf 1}_{\{\cdot\}}$ is indicator function and $|\!|\cdot |\!|$ is euclidean norm in $\mathbb{R}^d$ (for more information on Levy processes see \cite{B},\cite{Sa}).
\par For given Levy process $X$, the modelling of risky asset  can be made by the exponential process $S=(S_t)_{t\geq 0}$ with 
$$S_t = \,^{\top}\! ( \mathcal{E}(X^1)_t, \mathcal{E}(X^2)_t,\cdots,\mathcal{E}(X^d)_t)$$
where $\mathcal{E}(\cdot )$ is Dol\'{e}an-Dade exponential and $X^i$,$1\leq i\leq d,$ are the components of $X$.
 We recall that for each one-dimensional semi-martingale $X^i$,
\begin{equation*}
\mathcal{E}(X^i)_t=\exp\big\{X^i_t-\frac{1}{2}<X^{i,c}>_t\big\}\prod_{0\leq s\leq t} \exp\{-\Delta X^i_s\}(1+\Delta X^i_s)\,. 
\end{equation*}
Here  $<X^{i,c}>$ is quadratic variation of continuous martingale part of $X^i$ and $\Delta X^i$ are jumps of $X^i$ (see \cite{JSh} for more details).
\par Utility maximisation of exponential Levy models with single Levy process was considered in a number of articles (see for instance \cite{CV1}, \cite{CV2} and references there). However, the same questions in presence of illiquid assets in portfolio was not completely studied.
\par Utility maximisation in mentioned situation  was considered in a number of books and papers, see for instance \cite{BJ}, \cite{BFG},  \cite{Ca}, \cite{HeH}, \cite{MZ1}, \cite{MZ2}. Some explicit formulas for maximum of expected utility were obtained for Brownian motion models, where the incompleteness on the market comes from the non-traded asset (see \cite{HeH}, \cite{MZ1}, \cite{MZ2}). The formulas for maximum of expected utility in complete markets was derived in \cite{ABSch}. But the case of correlated Levy models with jumps was not considered up to now. 
\par To model dependent assets of Levy type, we denote by $X^{(1)}$ and $X^{(2)}$ two $d$-dimensional independent integrable Levy processes with generating triplets $(b_1,c_1, K _1)$ and $(b_2,c_2, K _2)$ respectively where $K_1$ and $K_2$ verify the condition of type \eqref{0}.  For two invertible matrix  $\rho _1$ and $\rho _2$ with real valued components, we introduce the process $X=(X_t)_{0\leq t\leq T}$ as a linear map
of $X^{(1)}$ and $X^{(2)}$, namely
$$X_t= \rho _1X^{(1)}_t+\rho _2X^{(2)}_t$$
We suppose that our two risky assets can be modelled by the processes $S^{(1)}=(S^{(1)}_t)_{0\leq t\leq T}$ and $S^{(2)}=(S^{(2)}_t)_{0\leq t\leq T'}$ with $T'> T$ and 
$$S^{(1)}_t= \,^{\top}\! ( \mathcal{E}(X^1)_t, \mathcal{E}(X^2)_t,\cdots,\mathcal{E}(X^d)_t)$$ and $$ S^{(2)}_t= \,^{\top}\! ( \mathcal{E}(X^{(2),1})_t, \mathcal{E}(X^{(2),2})_t,\cdots,\mathcal{E}(X^{(2),d})_t) .$$
To ensure that the components of $S^{(1)}$ and $S^{(2)}$ are positive, we assume, that for $1\leq i\leq d$,  the  jumps of $X^i$ and $X^{(2),i}$ verify: $\Delta X^i_t> ~-1,$ $\Delta X^{(2),i}_t>~-1$. Without loss of generality and up to now we assume that the interest rate $r$ of non-risky asset is equal to zero.
 \par In our setting,  the investor, which has two assets $S^{(1)}$ and $S^{(2)}$, can trade the first asset $S^{(1)}$ with maturity time $T$, but the second asset with maturity time $T'>T$, can not be traded
because of lack of liquidity or legal restrictions.
At the same time the investor has an European option $g(X^{(2)}_{T'})$ on risky asset  $S^{(2)}$, where $g$ is some non-negative real valued Borel function on $\mathbb{R}^d$. 
In such situation the investor, who would like to sell the option, would like also to evaluate  the corresponding maximal expected utility of the portfolio with option.
\par  Let us denote by $\Pi(\mathbb{F})$ the set of self-financing admissible strategies with respect to the filtration $\mathbb{F}$, generated by $X$. Then, for utility function $u$ and initial capital $x_0$, the optimal expected utility $U_T(x_0,0)$ related with  the first asset $S^{(1)}$  only, verify
$$U_T(x_0,0) = \sup_{\phi\in \Pi(\mathbb{F})} {\bf E} [u(x_0+\int_0^T\phi_s \cdot dS^{(1)}_s)]$$
and if we add the mentioned option, then the optimal expected utility will be equal to
$$U_T(x_0,g) = \sup_{\phi\in \Pi(\mathbb{G})} {\bf E} [u(x_0+\int_0^T\phi_s \cdot dS^{(1)}_s\,+\,g(X^{(2)}_{T'})]$$
where $\Pi(\mathbb{G})$ is a set of self-financing admissible strategies with respect to the enlarged filtration $\mathbb{G}= (\mathcal{G}_t)_{0\leq t \leq T}$ with, for $0\leq t<T$,
$$\mathcal{G}_t= \bigcap _{s>t}\mathcal{F}_s\otimes \sigma (X^{(2)}_{T'}) \hspace{0.5cm} \mbox{and}\hspace{0.5cm} 
 \mathcal{G}_T= \mathcal{F}_T\otimes \sigma (X^{(2)}_{T'}).$$
This approach coincide with so called utility maximisation with distortion
. In the case of Levy processes the distortion is $\delta =X^{(2)}_{T'}-X^{(2)}_T$,
and the information contained in $\mathcal{G}_T$ coincide with the one's of the $\sigma$- algebra
$\mathcal{F}^{(1)}_T\otimes\mathcal{F}^{(2)}_T\vee\sigma (X^{(2)}_{T'}-X^{(2)}_T)$,
i.e. with progressive filtration at time $T$ augmented by $\sigma$-algebra generated by distortion.

\par In this note we concentrate ourselves on non-complete market case
modelled by correlated exponential Lévy models.
 We recall that very often the utility maximisation analysis is carried out for the hyperbolic absolute risk utilities (in short HARA utilities). HARA utilities can be defined trough the coefficient of absolute risk aversion:
$$R(x) = - \frac{u''(x)}{u'(x)}$$
In HARA utility case,  $$R(x)=\frac{1}{A+Bx}$$ with $A$ and $B$ positive constants. The solutions of such differential equation for $u$ are known, 
and they are  logarithmic, power and exponential utilities  given below:
\begin{eqnarray*}
u(x) & =&\ln x, \,\,\mbox{with}\,\, x\in \mathbb{R}^{+,*},\\
u(x) & =& \frac{x^p}{p},\,\, \mbox{with}\,\,\,x\in \mathbb{R}^{+,*} \,\mbox{and}\,\ p\in(-\infty,0)\cup(0,1),\\
u(x) & =& 1-e^{-\gamma x},\ \mbox{with} \,\,x\in \mathbb{R}\, \mbox{and}\, \gamma>0.
\end{eqnarray*}
The problem of utility maximisation with option, when 
 $X$ and $X^{(2)}$ are semimartingales, 
was considered in \cite{EV}.  The method applied was based on enlargement of filtration, combined with the conditioning with respect to the variable $X^{(2)}_{T'}$ and, then, with dual approach. In dual approach we replace the problem of maximisation of expected utility by finding so-called f-divergence minimal martingale measure where 
$f$ is dual to $u$ function, namely
 \begin{equation}\label{FL}
 f(x)= \sup _{y\in \mathbb R} (u(y) -xy),
 \end{equation}
As known, if $u$ is logarithmic, then 
$$\hspace{0.5cm}f(x) = -\ln{x}-1,$$
if $u$ is power, then
$$f(x)= \frac{1-p}{p}x^{\frac{p}{p-1}},$$
and if $u$ is exponential,
$$\hspace{1cm}f(x) = 1-\frac{x}{\gamma}(1+\ln \gamma)+\frac{1}{\gamma}x\ln{x}.$$
 \par In Section \ref{s2}  we give, for convenience of the reader,  some results  about utility maximisation with option for semimartingale models. The main results of this section are the formulas for maximum of utility. These formulas contain the corresponding information quantities, like
 Kulback-Leibler information and Hellinger type integrals. In turn, these information quantities can be recovered
  from respective information processes.
\par In Section 3, we consider the exponential Levy models. More precisely, we verify the assumptions of Section 2 and we give the expressions for Girsanov parameters of f-divergence minimal conditional martingale measures. These expressions permit us to write the corresponding  information processes, and, then use the results of Section ~2.
\par  In Section \ref{s4} we give the expressions of the information quantities for  Black-Scholes models with correlated Brownian Motions.
\par In Section \ref{s5} we consider  Black-Scholes models with correlated Brownian Motions and jumps represented by Poisson process, in order to derive the  mentioned information quantities.
\end{section}

%%%%%%%%%%%%%%%%%%%%%%%%%%%%%%%%%%%%%%%%%%%%%%%%%%%%%%%%%%%%%%%%%%%%%%%%%%%%%%
\begin{section}{Some known results about utility maximisation with option for exponential semimartingale models.}\label{s2}%%%%%%
%%%%%%%%%%%%%%%%%%%%%%%%%%%%%%%%%%%%%%%%%%%%%%%%%%%%%%%%%%%%%%%%%%%%%%%%%%%%%%

\subsection{Modelling and assumptions}
\par We suppose that the process $X= (X_t)_{0\leq t\leq T}$ is given on canonical probability space $(\Omega , \mathcal{F}, P)$ with filtration $\mathbb{F}=(\mathcal{F}_t)_{0\leq t \leq T}$ satisfying usual properties. This process represents  stochastic logarithms of a $d$-dimensional liquid asset $S^{(1)}= (S^{(1)}_t)_{0\leq t \leq T}$ with
$$ S^{(1)}_t= \, ^{\top}\! ( \mathcal{E}(X^1)_t, \mathcal{E}(X^2)_t,\cdots,\mathcal{E}(X^d)_t)\,.$$
At the same time, we have also a $d$-dimensional semimartingale $X^{(2)}$, which represents stochastic logarithms of another risky, but illiquid asset. This illiquid asset, in turn, is represented in portfolio by European type option $g(X^{(2)}_{T'})$ where $g$ is a positive measurable function on $\mathbb{R}^d$ and $T'>~T$. 
\par To perform utility maximisation, we introduce a  product space $$(\Omega \times \mathbb{R}^d, \mathcal{F}\otimes \sigma (X^{(2)}_{T'}), P\times \alpha)$$ with $P$   "historical" law of $X$ and  $\alpha $  "historical" law of $X^{(2)}_{T'}$, endowed with enlarged filtration
$\mathbb{G}=(\mathcal{G}_t)_{0\leq T}$ with 
\begin{equation}
\mathcal{G}_t= \bigcap _{s>t}\mathcal{F}_s\otimes \sigma (X^{(2)}_{T'}) \hspace{0.5cm} \mbox{and}\hspace{0.5cm} 
 \mathcal{G}_T= \mathcal{F}_T\otimes \sigma (X^{(2)}_{T'}).
\label{6}
\end{equation}
We remark that $X$  remains a semimartingale on product space equipped
with filtration $\mathbb{G}$ since $X$ and $X^{(2)}$ are independent under the probability measure  $P\times \alpha$.
\par Now, we denote by $\mathbb{P}$ the law of the couple $( X, X^{(2)}_{T'})$ and by
$P^v$ the regular conditional law of $X$ given $\{X^{(2)}_{T'}=v\}$, i.e.
 for all $A\in \mathcal F$ and $v\in \mathbb{R}^d$
$$P^v(A) = \mathbb{P}(A\,|\,X^{(2)}_{T'}=v)\,.$$
To preserve semimartingale
property of $X$ under conditioning, we suppose that the following assumption  holds.
\asm \label{a1} For each $v\in\mathbb{R}^d$ the probability $P^v$ is locally absolutely continuous with respect to $P$, i.e
$$P^v\stackrel{loc}{\ll} P.$$
\rm
Under the Assumption 1 and according to \cite{J2} and \cite{JSh}, a semimartingale $X$ will remain a semimartingale under each  measure $P^v$, $v\in \mathbb{R}^d$.
Of course, the characteristics of a semimartingale $X$ under $P^v$ will be changed as it was proved in \cite{JSh} (cf. Theorem 3.24, p. 159).
\par For $0\leq t \leq T$, we denote by $P^v_t$ and $P_t$ the restrictions of the measures $P^v$ and $P$ on the $\sigma$-algebra $\mathcal{F}_t$. To avoid  measurability problems in semimartingale calculus depending on a parameter $v$ (cf. \cite{StY}), we need the optional versions of likelihood processes $( \frac{dP^v_t}{dP_t})_{0\leq t\leq T}$ with respect to the filtration $\mathbb{G}$. For that, we introduce  conditional distribution of  $X^{(2)}_{T'}$ given $\mathcal{F}_t$, i.e. 
\begin{equation*}
\alpha^t(\omega, dv)=\mathbb{P}(X^{(2)}_{T'}\in dv|\mathcal{F}_t)(\omega).
\end{equation*}
We make the following assumption
\asm \label{a2}The  conditional distributions of random variable $X^{(2)}_{T'}$ given  $\mathcal{F}_t$ are absolutely continuous with respect to its law, i.e.
$$\alpha^t\ll\alpha,\ \  \forall t\in\left]0,T\right].$$\rm
According to Jacod's lemma (see\cite{J1}), under the Assumption 2, there exists an optional version of density
process $( \frac{d\alpha ^{t}}{d\alpha})_{0\leq t\leq T}$.
\rem It should be noticed that the Assumption 2 can be replaced by the assumption that $\frac{dP^v_T}{dP_T}$ considered as a map of $(\omega, v)$ is $\mathcal F_T\otimes \mathcal{B}(\mathbb{R}^d)$- measurable. Then we can construct an optional version of density process using the results of \cite{StY}.\rm 
\par As it was mentioned, the next step consists to solve conditional utility maximisation problem using dual approach (see, for example, \cite{GR}). Let us denote by $f$ dual conjugate of utility function $u$.
Let $\mathcal{M}^v_T$ be the set of equivalent martingale measures on probability space $(\Omega , \mathcal{F}_T, P^v_T)$ for exponential semimartingale $S^{(1)}$ and let
\begin{equation*}
\mathcal{K}^v_T=\Big\{Q^v_T\in\mathcal{M}^v_T:\,{\bf E}_{P^v}\left|f\left(\frac{ dQ^v_T}{dP^v_T}\right)\right| < \infty \Big\}.
\end{equation*}
We recall that $Q^{v,*}_T$ is an equivalent f-divergence minimal martingale measure if 
$${\bf E}_{P^v} \big[ f(\frac{dQ^{v,*}_T}{dP^v_T})\big]= \min _{Q^v_T\in \mathcal{K}^v_T}{\bf E}_{P^v} \big[f(\frac{dQ^{v}_T}{dP^v_T} ) \big ]\,. $$ 
To use dual approach we introduce   the following assumption.
\asm \label{a3} For each $v\in \mathbb{R}^d$, there exists $f$-divergence minimal equivalent martingale measure $Q^{v, *}_T$, which  belongs to the set $\mathcal{K}_T^{v}$ and such that $z^*_T(v)= \frac{dQ^{v,*}_T}{dP^v_T}$ considered as a map of $(\omega, v)$, is $\mathcal F_T\otimes \mathcal{B}(\mathbb{R}^d)$- measurable and such that ${\bf E}_{P^v}f(z^*_T(v))$ is integrable in $v$ with respect to ~$\alpha$.\normalfont

\subsection{Existence of f-divergence minimal martingale measure}

\par  We recall  the results about the existence of global f-divergence minimal martingale measure.  For that we denote by $\mathbb{P}_T$ the restrictions of the measure $\mathbb{P}$ to the $\sigma-$algebra $\mathcal{G}_T$ and let $\mathcal{M}_T$ be the set of equivalent martingale measures for semimartingale $(S^{(1)}_t)_{0\leq t\leq T}$ considered as an application on probability space $(\Omega ^{(1)}_T\times \mathbb{R}^d, \mathcal{F}^{(1)}\otimes \mathcal{B}(\mathbb{R}^d), \mathbb{P}_T)$ with the filtration $\mathbb{G}$.
 Let
$$ \mathcal{K}_T = \{\mathbb{Q}_T\in \mathcal{M}_T \mid {\bf E}_{\mathbb{P}} |f(\frac{d\mathbb{Q}_T}{d\mathbb{P}_T})|< \infty\}\,.$$
We remark that $\mathcal{K}_T\neq \emptyset$. In fact, as Radon-Nikodym density of a measure $\mathbb{Q}_T$ with respect to $\mathbb{P}_T$,  we can take $z^*_T(v)$ from Assumption \ref{a3} with replacement of $v$ by  $X^{(2)}_{T'}$.  We recall that $\mathbb{Q}^*_T$ is $f$-divergence minimal measure if
$$\inf_{\mathbb{Q}_T\in \mathcal{K}_T} {\bf E}_{\mathbb{P}} f(\frac{d\mathbb{Q}_T}{d\mathbb{P}_T})={\bf E}_{\mathbb{P}} f(\frac{d\mathbb{Q}^*_T}{d\mathbb{P}_T})\,.$$
%%%%%%%%%%%%%%%%%%%%%%%%%%%%%%%%%%%%%%%%%%%%%%%%%%%%%%%%%%%%%%%%%%%%%%%%%%%%%%%
\thm \label{t1} \rm (cf. \cite{EV})\it 
%%%%%%%%%%%%%%%%%%%%%%%%%%%%%%%%%%%%%%%%%%%%%%%%%%%%%%%%%%%%%%%%%%%%%%%%%%%%%%
Under the Assumptions 1,2,3 there exists $\mathbb{Q}^*_T\in \mathcal{K}_T$ which is $f$-divergence minimal martingale measure and
\begin{equation}
U_T(x_0,g)=\int_{\mathbb{R}^d}{\bf E}_{P^v}\left[u\left(-f'\left(\lambda_g(v)z^{*}_T(v)\right)\right)\right]d\alpha(v),
\label{64}
\end{equation}
where $\lambda_g(v)$ is a solution of the equation
\begin{equation}\label{62}
-{\bf E}_{P^v}[f'(\lambda _g(v)\, z^*_T(v))] = x_0 +g(v) \,.
\end{equation}

\normalfont

\subsection{Conditional information quantities and maximal expected utility}

\par Let us assume the existence of $f$-divergence minimal martingale measure $Q^{v, *}_T\in\mathcal{K}^v_T$ and denote
$$z_T^{*}(v)= \frac{Q^{v, *}_T}{P^v_T},\,\,\,p^v_T=\frac{dP^v_T}{dP_T}\,.$$ Now,
we introduce three important quantities related with $P^v_T$ and $Q^{v, *}_T$ namely the entropy of $P_T^{v}$ with respect to $Q^{v, *}_T$,
\begin{equation*}
{\bf{I}}(P^{v}_T|Q^{v,*}_T)=-{\bf E}_{P^v}\left[\ln{z_T^{*}(v)}\right]=-{\bf E}_{P}\left[p^v_T\,\ln{z_T^{*}(v)}\right],
\end{equation*}
then, the entropy of $Q^{v, *}_T$ with respect to $P^{v}_T$ or Kulback-Leibler information
\begin{equation*}
{\bf{I}}(Q^{v, *}_T|P^{v}_T)={\bf E}_{P^v}\left[z_T^{*}(v)\ln{z_T^{*}(v)}\right]={\bf E}_{P}\left[p^v_T\,z_T^{*}(v)\ln{z_T^{*}(v)}\right],
\end{equation*}
and also  Hellinger type integrals 
\begin{equation*}
{\bf H}^{(q), *}_T(v)={\bf E}_{P^v}\left[(z_T^{*}(v))^q\right]={\bf E}_{P}\left[p^v_T\,(z_T^{*}(v))^q\right], 
\end{equation*}
where $q=\frac{p}{p-1}$ with $ p<1$.
\par In the following theorem we give the expressions of the maximal expected utility involving relative entropies and Hellinger-type integrals.
%%%%%%%%%%%%%%%%%%%%%%%%%%%%%%%%%%%%%%%%%%%%%%%%%%%%%%%%%%%%%%%%%%%%%%%%%%%%%
\thm \label{t2} \rm (cf. \cite{EV})\it
%%%%%%%%%%%%%%%%%%%%%%%%%%%%%%%%%%%%%%%%%%%%%%%%%%%%%%%%%%%%%%%%%%%%%%%%%%%%%%
Under the Assumptions \ref{a1}, \ref{a2}, \ref{a3}, there exist a $\mathcal{B}(\mathbb{R}^d)$-measurable versions of the information quantities. Moreover,  we have the following expressions for $U_T(x_0,g):$

$(i)$ If $u(x)=\ln{x}$  then 
\begin{equation}
U_T(x_0,g)=\int_{\mathbb{R}^d}[\,\ln(x_0+g(v))+{\bf{I}}(P^{v}_T|Q^{v, *}_T)\,]d\alpha(v)\,.
\label{72a}
\end{equation}

$(ii)$ If $u(x)=\frac{x^p}{p}$ with $p<1, p\neq 0$ then
\begin{equation}
U_T(x_0,g)=\frac{1}{p}\int_{\mathbb{R}^d}(x_0+g(v))^p\left({\bf H}^{(q), *}_T(v)\right)^{1-p}d\alpha(v)\,.
\label{73a}
\end{equation}

$(iii)$ If $u(x)=1-e^{-\gamma x}$ with $\gamma>0$  then
\begin{equation}
U_T(x_0,g)=1-\int_{\mathbb{R}^d}\,\exp \{-[\,\gamma (x_0+g(v))+{\bf I}(Q^{v, *}_T|P^{v}_T)\,]\}\,d\alpha(v)\,.
\label{74a}
\end{equation}

\normalfont

\subsection{Conditional information processes and conditional information quantities}

\par In this subsection we recall  that the conditional information quantities can be recovered from conditional information processes.
To simplify the expression for information processes
we suppose during this subsection that the process $X$ is quasi-left continuous. We recall that  $(P,\mathbb F)$-semimartingale $X$ is a quasi-left continuous, if for any predictable stopping time $\tau$, the jump $\Delta X_{\tau}=0$ ($P$-a.s.) on the set  $\{\tau <\infty \}$.
We remark that since $P^v\stackrel{loc}{\ll}P$, $(P^v,\mathbb{F})$ semi-martingale $X$ will be also quasi-left continuous.
\par Let us denote by $\beta^{v, *}$ and $Y^{v, *}$ two $(P^v,\mathbb F)$-predictable processes known as Girsanov parameters for the change of measure $P^v$ into $Q^{v, *}$ such that: $\forall t\geq 0$ and $P^v$-a.s.
\begin{equation*}
\int_0^t\int_{\mathbb{R}^d}|\!|l(x)|\!|\,|(Y_s^{v,*}(x)-1)|\,\nu^v(ds,dx)<\infty,
\end{equation*}
and
$$ \int_0^t|\!|c_s\,\beta_s^{v,*})|\!|ds<\infty ,\,\, \int_0^t\, ^{\top}\beta_s^{v,*} c_s \beta_s^{v,*} ds<\infty \,.$$

Here $\nu^{v}$  stands for the compensator of the jump measure of $X$ with respect to $(P^v,\mathbb F)$, $l$ is the truncation function and $c$ is the density of the predictable variation of continuous martingale part of $X$, w.r.t. Lebesgue measure.
\par In the case of logarithmic utility we consider the entropy ${\bf{I}}(P^{v}_T\,|\,Q^{v, *}_t)$ and also the corresponding information process  
$\mathcal{I}^{*}(v)=(\mathcal{I}^{*}_t(v))_{t\in\left[0,T\right]}$ with
\begin{equation}
\hspace{-5cm}\mathcal{I}^{*}_t(u)=  \frac{1}{2}\int_0^t\,^{\top}\beta_s^{v,*} c_s \beta_s^{v,*} ds 
\label{71}
\end{equation}
$$\hspace{3cm}-\int_0^t\int_{\mathbb{R}^d}\left[\ln(Y_s^{v, *}(x))-Y_s^{v, *}(x)+1\right]\nu^v(ds,dx).$$
%%%%%%%%%%%%%%%%%%%%%%%%%%%%%%%%%%%%%%%%%%%%%%%%%%%%%%%%%%%%%%%%%%%%%%%%%%
\prop \label{p1}
%%%%%%%%%%%%%%%%%%%%%%%%%%%%%%%%%%%%%%%%%%%%%%%%%%%%%%%%%%%%%%%%%%%%%%%%%%
Let $Q^{v,*}_T\in \mathcal{K}^v_T$. Then  the corresponding relative entropy  is well-defined and
\begin{equation}
{\bf{I}}(P^{v}_T\,|\,Q^{v, *}_T)={\bf E}_{P^v}\mathcal{I}^{*}_T(v).
 \label{76}
 \end{equation}\rm

\par In the case of exponential utility we consider  Kullback-Leiber information ${\bf{I}}(\,Q^{v,*}_T\,|\,P^{v}_T\,)$ and we introduce the corresponding Kullback-Leiber process ${I}^{*}(v)=({I}^{*}_t(v))_{t\in\left[0,T\right]}$ with
\begin{equation}\hspace{-5cm}{I}^{*}_t(v)=\frac{1}{2}\int_0^t\,^{\top}\beta_s^{v,*} c_s \beta_s^{v,*} ds 
\label{83}
\end{equation}
$$\hspace{3cm}+\int_0^t\int_{\mathbb{R}^d}\left[Y_s^{v, *}(x)\ln(Y_s^{v, *}(x))-Y_s^{v, *}(x)+1\right]\nu^v(ds,dx).$$
%%%%%%%%%%%%%%%%%%%%%%%%%%%%%%%%%%%%%%%%%%%%%%%%%%%%%%%%%%%%%%%%%%%%%%%%%%%%%%%
\prop \label{p2} 
%%%%%%%%%%%%%%%%%%%%%%%%%%%%%%%%%%%%%%%%%%%%%%%%%%%%%%%%%%%%%%%%%%%%%%%%%%%%%%

Let $Q^{v,*}_T\in \mathcal{K}^v_T$. Then, the corresponding Kullback-Leibler information is well defined and
\begin{equation}
{\bf{I}}(\,Q^{v, *}_T\,|\,P^{v}_T\,)={\bf E}_{P^v}\left[\int_0^T\, z_{s-}^{*}(v)\,d{I}^{*}_s(v)\right] = {\bf E}_{Q^{v,*}}\left(\,{I}^{*}_T(v)\,\right)\,. 
 \label{83a}
 \end{equation}
\normalfont 

\par For the case of power utility we consider Hellinger types integrals 
$${\bf H}^{(q), *}_T(v)={\bf E}_{P^v}\left[(z_T^{*}(v))^q\right],$$
where $ q=\frac{p}{p-1}<1$.
 \par We introduce the corresponding predictable process called Hellinger type process $h^{(q), *}(v)=(h^{(q), *}_t(v))_{t\in\left[0,T\right]}$
\begin{equation}\label{88}
\hspace{-3cm}h^{(q), *}_t(v)=\frac{1}{2}q(1-q)\int_0^t\,^{\top}\beta_s^{v,*} c_s \beta_s^{v,*} ds 
\end{equation}
$$\hspace{3cm} - \int_0^t\int_{\mathbb{R}^d}\left[\left(Y_s^{v, *}(x)\right)^{q}-q(Y_s^{v, *}(x)-1) -1\right]\nu ^v(ds,dx)\,.$$

%%%%%%%%%%%%%%%%%%%%%%%%%%%%%%%%%%%%%%%%%%%%%%%%%%%%%%%%%%%%%%%%%%%%%%%%%%%%%%%%%%%%%%%%%
\prop \label{p3} Let $Q^{v,*}_T\in \mathcal{K}^v_T$. Then respective Hellinger type integral  ${\bf H}^{(q), *}_T(v)$  is well defined and
\begin{equation}
{\bf H}^{(q), *}_T(v)=1-{\bf E}_{P^v}\left[\int_0^T(z_{s-}^{*}(v))^q\,dh_s^{(q),*}(v)\right]
\end{equation}\label{hel1}
or, in the terms of the stochastic exponential,
\begin{equation}
{\bf H}^{(q), *}_T(v)={\bf E}_{R^v}\left[\mathcal{E}\left(-h^{(q), *}(v)\right)_T\right]
\label{hel2}
\end{equation}
where $R^v$ is some locally absolutely continuous w.r.t. $P^v$ measure.
%%%%%%%%%%%%%%%%%%%%%%%%%%%%%%%%%%%%%%%%%%%%%%%%%%%%%%%%%%%%%%%%%%%%%%%%%%%%%%%%%%%%%%%%%
\normalfont

\end{section}

%%%%%%%%%%%%%%%%%%%%%%%%%%%%%%%%%%%%%%%%%%%%%%%%%%%%%%%%%%%%%%%%%%%%%%%%%%%%
\begin{section}{ Utility maximisation with option for exponential Lévy models}\label{s3}
%%%%%%%%%%%%%%%%%%%%%%%%%%%%%%%%%%%%%%%%%%%%%%%%%%%%%%%%%%%%%%%%%%%%%%%%%%%
We begin with some basic notations  for the exponential Lévy models
involved in the utility maximisation calculus.

\subsection{Description of the model}
 Let $X^{(1)}=(X^{(1)}_t)_{0\leq t\leq T}$ and $X^{(2)}=(X^{(2)}_t)_{0\leq t\leq T'}$ be two independent $d$-dimensional Levy processes starting from zero with generating triplets $(b_1,c_1,K _1)$ and $(b_2,c_2,K _2)$ respectively. Each process is given on its own filtered canonical probability space $(\Omega ^{(1)}, \mathcal{F}^{(1)}, \mathbb{F}^{(1)},P^{(1)})$ and $(\Omega ^{(2)}, \mathcal{F}^{(2)}, \mathbb{F}^{(2)},P^{(2)})$ respectively where $\mathbb{F}^{(1)}= (\mathcal{F}^{(1)}_t)_{0\leq t \leq T}$ and
$\mathbb{F}^{(2)}= (\mathcal{F}^{(2)}_t)_{0\leq t \leq T'}$ are the corresponding filtrations verifying usual properties.
 Let $X=(X_t)_{0\leq t \leq T}$ be the linear map of the processes $X^{(1)}$ and $X^{(2)}$, namely, for $t\in [0,T]$
\begin{equation}\label{eq}
X_t= \rho _1 \,X^{(1)}_t + \rho _2 \,X^{(2)}_t
\end{equation}
involving non-random invertible matrices $\rho _1$ and $\rho _2$.
As it was mentioned, the process $X$ is considered  on canonical probability  space $(\Omega , \mathcal{F}, \mathbb{F}, P)$  with  filtration $\mathbb{F}=(\mathcal{F}_t)_{0\leq t\leq T}$ which satisfy usual properties.
\par We introduce also the enlarged space $(\Omega \times \mathbb{R}^d, \mathcal{F}\otimes \sigma(X^{(2)}_{T'}), \mathbb{G})$, corresponding to the couple $(X, X^{(2)}_{T'})$ with enlarged filtration $\mathbb{G}=(\mathcal{G}_t)_{\leq t \leq T}$ where for $0\leq t<T$
$$\mathcal{G}_t= \bigcap _{s>t}\mathcal{F}_s\otimes \sigma (X^{(2)}_{T'}) \hspace{0.5cm} \mbox{and}\hspace{0.5cm} 
 \mathcal{G}_T= \mathcal{F}_T\otimes \sigma (X^{(2)}_{T'}).$$
 We remark that on the space $(\Omega , \mathcal{F}, P)$ the process $X$, is, evidently, a Levy process. Now, if we equip $(\Omega \times \mathbb{R}^d, \mathcal{F}\otimes \mathcal{B}(\mathbb{R}^d), \mathbb{G})$
with the probability $P\times \alpha$, where $\alpha$ is the law of $X^{(2)}_{T'}$ , then the process $X$ will remain a Levy process with the same triplet.
We recall that, as before, we use the notation $\mathbb{P}$ for joint law of $(X, X^{(2)}_{T'})$ and
$P^v$ for conditional law of $X$ given $X^{(2)}_{T'}=v$.

\subsection{Assumptions 1 and 2}
In this subsection we show that the Assumptions 1 and 2 of Section 2 will be verified under the following hypothesis on Lévy processes.
\par {\bf Hypothesis}\,\it H1:  The processes $X$ and  $X^{(2)}$ are integrable on $[0,T]$ and $[0,T']$ respectively.\rm

\par{\bf Hypothesis}\,\it H2:  The process $(X, X^{(2)})$ has  a transition density w.r.t. a product  $\eta=\eta _1\times \eta _2$ of two $\sigma$-finite measures $\eta _1$ and $\eta _2$  on $\mathbb{R}^d$,    and the marginal transition densities $f$ and $f^{(2)}$  of $X$ and $X^{(2)}$ w.r.t. $\eta$ and $\eta _2$ respectively,  are strictly positive.\rm

\rem \rm It should be noticed that in the case when $\eta _1$ and $\eta _2$ are Lebesgue  measures, the Hypothesis 
\it H2 is equivalent to the existence of marginal strictly positive transition densities $f_1$ and $f_2$ of the processes of $X^{(1)}$ and $X^{(2)}$. This fact follows from the independence of $X^{(1)}$ and $X^{(2)}$.\\

\prop \label{l2} Under hypotheses $(H1)$ and $(H2)$,  the Assumptions 1 and 2 are satisfied and there exists  a function $F_v:[T'-T, T']\times \mathbb{R}^d\rightarrow \mathbb{R}^+$ depending on a parameter $v\in \mathbb{R}^d$, such that
\begin{equation}\label{c1}
\frac{d\alpha ^t}{d\alpha}(v)= \frac{F_v(T'-t,X_t)}{F_v(T',0)}
\end{equation}
Moreover,
$$\frac{d\alpha ^t}{d\alpha} = \mathcal{E} (M)_t$$
with  $M = (M_t)_{0\leq t\leq T}$ which is a $(P, \mathbb{F})$- martingale such that 
$$M_t= \int^t_0\,^{\top}\beta_s^{v,P}\, X_s^{c}+ \int^t_0\int _{\mathbb{R}^d} l(x) ( Y_s^{v,P}(x)-1) d K (x) ds $$
where $(\beta^{v,P}, Y^{v,P})$ are the Girsanov parameters for the  change the measure $P$ into $P^v$, and $K$ is Levy measure of $X$.
\par If $F_v\in C^{1,2}_b([T'-T, T']\times \mathbb{R}^d)$ and $c$ is invertible, then the mentioned Girsanov parameters $(\beta^{v,P}, Y^{v,P})$ can be calculated by the following formulas:
$$^{\top}\beta_s^{v,P} =  \left(\frac{\partial \ln F_v}{\partial x_1}(T'-s, X_{s-}), \cdots , \frac{\partial \ln F_v}{\partial x_d}(T'-s, X_{s-})\right)$$
and for $x\in \mathbb{R}^d\setminus\{0\}$
$$Y_s^{v,P}(x) = \frac{F_v(T-s, X_{s-}+x)}{F_v(T'-s,X_{s-})}\,.$$

\pf \rm  Conditionally to $X^{(2)}_{T'}=v$, the process $X$ is distributed as $\rho _1X^{(1)} +\rho _2 V^{(2)}$ where  $V^{(2)}$ is a Levy bridge of $X^{(2)}$ starting at $(0,0)$ and ending at $(v,T')$. Under the hypothesis (H2) and according to \cite{GVV}, the law of $(V^{(2)}_t)_{0\leq t \leq T}$ is absolutely continuous w.r.t.
the law of $(X^{(2)}_t)_{0\leq t \leq T}$ and
\begin{equation}\label{abs}
 \frac{dP_{V^{(2)}}}{dP_{X^{(2)}}}(T,v) = \frac{f^{(2)}(T'-T,v-X^{(2)}_T)}{f^{(2)}(T',v)}\,.
\end{equation}
Since the process $X^{(1)}$ is independent from $X^{(2)}$ and also from $V^{(2)}$,  the conditional distributions of $X$ given $X^{(2)}$  and the conditional distributions of $X$ given $V^{(2)}$  coincide as maps, under the measure $P$. Let us denote this map by $q(A,x),\, A\in \mathcal{F}_T, x\in \mathbb{R}^d$.
Then,
$$P(A)= \int_{\mathbb{R}^d} P(\rho _1\,X^{(1)}+\rho _2\,X^{(2)}\in A\,|\, X^{(2)}=x) dP_{X^{(2)}}(x) = \int_{\mathbb{R}^d} q(A,x)dP_{X^{(2)}}(x)$$
and
$$\hspace{-3cm}P^{v}(A) =  P(\rho _1\,X^{(1)}+\rho _2\,V^{(2)}\in A)$$ 
$$\hspace{3cm}= \int_{\mathbb{R}^d} P(\rho _1\,X^{(1)}+\rho _2\,V^{(2)}\in A\,|\, V^{(2)}=x)\,dP_{V^{(2)}}(x)$$ 
$$ \hspace{5cm}= \int_{\mathbb{R}^d} q(A,x) \,\frac{dP_{V^{(2)}}}{dP_{X^{(2)}}}(T,v)\,dP_{X^{(2)}}(x)\,.$$
Finally, if $P(A)=0$ then $q(A,x)=0$\, ($P_{X^{(2)}}$-a.s.) and, hence $P^{v}(A)=~0$.
Hence, the  Assumption 1 is verified.
\par 
The Assumption 1 and Bayes formula for conditional densities gives us: 
$$P(X^{(2)}_{T'}\in dv\,|\, X_t=y)=\frac{P(X^{(2)}_{T'}\in dv, X_t \in dy)}{P(X_t\in dy)}=$$
$$\frac{P( X_t \in dy\,|\,X^{(2)}_{T'}=v)\,P( X^{(2)}_{T'}\in dv)}{P(X_t\in dy)}\,.$$
This means that the Assumption 2 is verified.
Using Markov property we write:
$$\alpha ^t(dv) = P( X^{(2)}_{T'}\in dv\,|\, \mathcal{F}_t)=
P( X^{(2)}_{T'}\in dv\,|\,X_t)=$$
$$P( X^{(2)}_{T'}-X_t^{(2)}+X_t^{(2)}\in dv\,|\, X_t)= P( \tilde{X}^{(2)}_{T'-t}+X_t^{(2)}\in dv\,|\, X_t) $$
where $\tilde{X}^{(2)}$ is a process, which is independent from $X^{(1)}$ and $X^{(2)}$, and is distributed as $X^{(2)}$. Then, we see that $\alpha ^{t}(dv)$ is a function of $T'-t$, $X_t$ and the parameter $v$, denoted $F_v(T'-t, X_t)$. At the same time 
$\alpha^{0}(dv) = \alpha (dv) = F_v(T',0)$ since $\mathcal{F}_0=\{\emptyset, \Omega\}$. It gives us \eqref{c1}.
\par Now, we use Ito formula to obtain that
$$ F_v(T'-t, X_t)= F_v(T',0)$$
$$ - \int ^t_0 \frac{\partial F_v}{\partial s}( T'-s, X_{s-}) \,ds+ \sum_{i=1}^d\int ^t_0 \frac{\partial F_v}{\partial x_i}( T'-s, X_{s-}) \,dX^i_s$$
$$ + \frac{1}{2} \sum_{i=1}^d \sum_{j=1}^d \int ^t_0 \frac{\partial^2 F_v}{\partial x_i \partial x_j}( T'-s, X_{s-}) \,d<X^{i,c}, X^{j,c}>_s$$
$$+ \sum _{0< s \leq t} F_v( T'-s, X_{s}) -F_v( T'-s, X_{s-})
-  \sum_{i=1}^d\frac{\partial F_v}{\partial x_i}( T'-s, X_{s-}) \,\Delta X^i_s\,.$$
\par Under the conditions $P^v_t <\!\!< P_t$ and $\alpha ^t <\!\!< \alpha$ for $ t\in ]0,T],$ we know from Jacod's lemma (cf.\cite{J1}) that 
$\frac{d\alpha^t}{d\alpha}_{0\leq t\leq T}$ is a $(P, \mathbb{F})$ martingale.
Let us put for $t\in[0,T]$, $p^v_t=\displaystyle\frac{d\alpha^t}{d\alpha}(v)$.
Then, we  divide the above expression for $F_v(T'-t, X_t)$ by $F_v(T',0)$ and we identify its continuous martingale part. We get that 
$$p^{v,c}_t= \frac{1}{F_v(T',0)} \sum_{i=1}^d \int ^t_0 \frac{\partial F_v}{dx_i}( T'-s, X_{s-}) \,dX^{i,c}_s\,$$
and, hence,
 $$<p^{v,c},X^{j,c}>_t= \frac{1}{F_v(T',0)}\sum_{i=1}^d\int_0^t c_{i,j} \frac{\partial F_v}{dx_i}( T'-s, X_{s-}) \,ds\,.$$
In addition, according to Girsanov theorem,
 $$<p^{v,c},X^{j,c}>_t= \sum_{i=1}^d \int_0^t\,c_{i,j}(\beta^{v,P}_s)^{i}p^{v}_{s-}\,ds.$$
 Since $c$ is invertible, this implies the formula for $\beta^{v,P}_t$.\\
Now, we compute the jumps of $p^v$:
$$\Delta p^v_t = \frac{F_v( T'-t, X_{t}) -F_v( T'-t, X_{t-})}{F_v(T',0)}$$
and
$$\frac{\Delta p^v_t}{p^v_{t-}}=  \frac{F_v( T'-t, X_{t-}+\Delta X_t)}{F_v( T'-t, X_{t-})}-1 \,.$$
Then, according to the Theorem 3.24,p.159, Chapter 3 in \cite{JSh}
$$ Y^{v,P}_t= \frac{F_v( T'-t, X_{t-}+x))}{F_v( T'-t, X_{t-})}$$
and the proposition is proved.
$\Box$

\subsection{Conditional locally equivalent martingale measures}
\par The main difficulty related with the application of the results of Section 2 is the verification of the Assumption 3. The first step for this verification, consists in the complete description of the set of conditional locally equivalent martingale measures. This step can be done by use of semimartingale calculus.
\par We recall that the process $X$ is defined by \eqref{eq}. As before we denote by $(\beta^{v,P},Y^{v,P})$  the Girsanov parameters for the change of the measure $P$ into $P^{v}$. We denote also  by $\mathcal{M}(P^v)$ the set of locally equivalent to $P^v$ martingale measures $Q^v$. We denote by $(\beta^v, Y^v)$ the Girsanov parameters for the change of the measure $P^v$ into $Q^v$. We notice that $X$ is $(P^v, \mathbb{F})$- semimartingale, and hence, $(Q^v, \mathbb{F})$ semimartingale.  In the following proposition we give the triplet of predictable characteristics of $X$ w.r.t. $Q^v$.
\prop \label{pl1}
The triplet of predictable characteristics $( B ^v, C ^v, \nu ^v )$ of $X$ with respect to $(Q^v, \mathbb{F})$ is given by the expressions:
$$ \begin{array}{l}
B^v_t = (\rho _1b_1+ \rho _2 b_2)t + \rho_2 c_2 \int^t_0 \beta^{v,P}_s ds\\\\
 \hspace{0.5cm}+ \rho_2\int ^t_0 \int _{\mathbb{R}^d} \stackrel{\rightarrow}{l_2(x)} ( Y^{v,P}_s(\rho _2^{-1}x) -1) (K_2\circ\rho _2^{-1})(dx)ds\\\\
\hspace{0.5cm}+ \int ^t_0  \int _{\mathbb{R}^d} \stackrel{\rightarrow}{l(x)} (Y^v_s(x) -1)  K_s^{v,P}(dx) ds + ( \rho _1 c_1 \,^{\top}\rho _1 + \rho_2 c_2 \,^{\top}\rho _2 ) \int ^t_0 \beta^v_s ds,\\\\
C^v_t= ( \rho _1 c_1 \,^{\top}\rho _1 + \rho_2 c_2 \,^{\top}\rho _2 )t,\\\\
d\nu ^v(x,s) = Y^v_s \, K_s^{v,P}(dx) ds,
\end{array}$$
where $K_s^{v,P}(dx) = \,(K _1\circ\rho _1^{-1})(dx) + \,Y^{v,P}_s(\rho _2^{-1}x)\, (K_2\circ\rho _2^{-1})(dx)$.
Moreover, an equivalent to $P^v_T$ martingale measure $Q^v_T$  satisfy : for $s\in [0,T]$
\begin{equation}
\label{mart}
\rho _1b_1+ \rho _2 b_2 +  \rho _2 c_2 \beta^{v,P}_s  + \rho _2\int _{\mathbb{R}^d} \stackrel{\rightarrow}{l_2(x)} ( Y^{v,P}_s(\rho _2^{-1}x) -1) (K _2\circ\rho _2^{-1})(dx)
\end{equation}
$$\hspace{1.7cm}+   \int _{\mathbb{R}^d} \stackrel{\rightarrow}{l(x)} (Y^v_s(x) -1)  K_s ^{v,P}(dx) + (  \rho _1 c_1 \,^{\top}\rho _1 + \rho_2 c_2 \,^{\top}\rho _2 )  \beta^v_s =0. $$ 

\pf \rm We use Girsanov theorem for successive change of the measures: $P\rightarrow P^v \rightarrow Q^v$. For that we write first a semimartingale decomposition of $X$:
$$ X_t= B_t+ X^{c}_t+ \int ^t_0 \int _{\mathbb{R}^d}\, \stackrel{\rightarrow}{l(x)}(\mu _X(dx, ds)-\nu_X(dx , ds))+\sum_{s\leq t}(\Delta X_s - l(\Delta X_s))$$ 
Here $B$ is the drift part of semilmartingale decomposition, $X^c$ is its continuous martingale part, $\mu _X$ and $\nu _X$ are the measure of jumps and its compensator, and 
$l$ is the truncation function, $l(x) = x{\bf 1}_{\{|\!|x|\!|\}},\, x\in \mathbb{R}^d$.
It should be noticed that the integral on $\mathbb{R}^d$ in previous expression is taken in component by component way, namely for each $x\in \mathbb{R}^d$ and $l(x) = \,^{\top}(l^1(x), \cdots , l^d(x))$ the integral
$$\int ^t_0 \int _{\mathbb{R}^d} \stackrel{\rightarrow}{l(x)}(\mu _{X}(dx, ds)-\nu _X(dx, ds))$$ 
is a vector with components
$$\int ^t_0 \int _{\mathbb{R}^d}l^i(x)(\mu _{X}(dx, ds)-\nu _X(dx, ds))$$
where $1\leq i\leq d$. We use the notation  $\stackrel{\rightarrow}{l(x)}$ to underline this particular integration.
\par At the same time we write a semi-martingale decompositions of the processes $X^{(1)}$ and $X^{(2)}$:
$$ \hspace{-1cm}X^{(1)}_t= b_1t+ X^{(1),c}_t+ \int ^t_0 \int _{\mathbb{R}^d}\, \stackrel{\rightarrow}{l_1(x)}(\mu _{X^{(1)}}(dx, ds)-K_1(dx)ds)$$ $$\hspace{7cm}+\sum_{s\leq t}(\Delta X^{(1)}_s - l_1(\Delta X^{(1)}_s))$$
$$ \hspace{-1cm}X^{(2)}_t= b_2t+ X^{(2),c}_t+ \int ^t_0 \int _{\mathbb{R}^d} \stackrel{\rightarrow}{l_2(x)}(\mu _{X^{(2)}}(dx, ds)-K_2(dx)ds)$$ $$\hspace{7cm}+\sum_{s\leq t}(\Delta X^{(1)}_s - l_2(\Delta X^{(2)}_s))$$
with truncation functions $l_1(x) = x{\bf 1}_{\{|\!|\rho _1 x|\!|\leq 1\}}$ and 
$l_2(x) = x{\bf 1}_{\{|\!|\rho _2 x|\!|\leq 1\}}$ respectively.
 \par We compare now the linear combination of the canonical decompositions of the  processes $X^{(1)}$ and $X^{(2)}$ given above  with canonical decomposition of $X$.
We can  easily identify a drift part of $X$, which  is $(\rho _1 b_1 + \rho _2 b_2)t, \, t\geq 0$, and a continuous martingale part of $X$, which is equal to $\rho _1X^{(1),c}+ \rho _2X^{(2),c}$. Since $X^{(1),c}_t$ and $X^{(2),c}_t$ are independent with quadratic variations $c_1t,\,t\geq 0$ and $c_2t,\,t\geq 0$, the quadratic variation of continuous martingale part of $X$ is equal to $(\rho _1 c_1 \,^{\top}\rho _1 + \rho_2 c_2 \,^{\top}\rho _2)t,\,t\geq 0$. 
\par For jump-part we write the measure of jumps of the process $X$:
$$ \mu _X(\omega, dt, dx) = \sum _{s}{\bf 1}_{\{\Delta X_s(\omega)\neq 0 \}}\delta_{\{(s,\Delta X_s(\omega)) \}}(dt,dx)$$
where $\delta$ is Dirac delta-function in $\mathbb{R}^{d+1}$.
In addition,
$$\Delta X= \rho _1 \Delta X^{(1)}+ \rho _2 \Delta X^{(2)},\,\,\,\,l(\Delta X)= \rho _1 l_1(\Delta X^{(1)})+ \rho _2 l_2(\Delta X^{(2)}).$$
We know that two independent Levy processes can not jump at the same time. In fact, the jumps of Levy processes are totally inaccessible stopping times. If we suppose that the jumps of the processes $X^{(1)}$ and $X^{(2)}$ happen at the times $\tau _1$ and $\tau _2$ with $\tau_1= \tau _2$ (P-a.s.) then 
for all $A\in \mathbb{R}^d$
$$P(\{\tau _1\in A\}\cap \{\tau _2\in A\})= P(\{\tau _1\in A\})=P^2(\{\tau _1\in A\})\,.$$
Then, $P(\{\tau _1\in A\})=0$ or 1, and the law of $\tau _1$ can be only Dirac measure. Then, there exists $t_0\in \mathbb{R}^+$ such that $P(\tau _1= t_0)=1$, but this contradicts with the fact that $\tau _1$ is inaccessible. This fact gives us that $P-a.s.$
$$\{\Delta X_s(\omega)\neq 0 \}= \{\rho _1\Delta X^{(1)}_s(\omega)\neq 0 \}\cup \{\rho _2\Delta X^{(2)}_s(\omega)\neq 0 \} $$
$$=  \{\rho _1\Delta X^{(1)}_s(\omega)\neq 0 \}\cap \{\Delta X^{(2)}_s(\omega)= 0 \} \cup \{\Delta X^{(1)}_s(\omega)= 0 \}  \{\rho _2\Delta X^{(2)}_s(\omega)\neq 0 \}\,.$$
As a consequence,\\
$$\mu _X(\omega, dt, dx)=$$
$$\sum _{s}{\bf 1}_{\{\Delta X^{(1)}_s(\omega)\neq 0 \}}\delta_{\{(s,\rho _1\Delta X^{(1)}_s(\omega)) \}}(dt,dx)+\sum _{s}{\bf 1}_{\{\Delta X^{(2)}_s(\omega)\neq 0 \}}\delta_{\{(s,\rho _2\Delta X^{(2)}_s(\omega)) \}}(dt,dx)$$
$$= \mu _{\rho _1X^{(1)}}(\omega, dt, dx) + \mu _{\rho _2X^{(2)}}(\omega, dt, dx)\,.$$
Now, the processes $\rho _1 X^{(1)}$ and $\rho _2 X^{(2)}$ are Levy processes with Levy measures $K_1(\rho _1^{-1}A)$ and $K_2(\rho _2^{-1}A)$ respectively where $A\in \mathcal{B}(\mathbb{R}^d)$. As a consequence, the triplet of predictable characteristics $(B,C,\nu)$ of $X$ is given by:
$$\begin{array}{l}
B_t= (\rho _1 b_1 + \rho _2 b_2)t,\\

C_t=  (\rho _1 c_1 \,^{\top}\rho _1 + \rho_2 c_2 \,^{\top}\rho _2)t,\\

d\nu (x,t) = \left((K_1\circ\rho _1^{-1})(dx) + (K_2\circ\rho _2^{-1})(dx)\right)dt.
\end{array}$$
\par Next , we write the triplet $( B^{V}, C^{V}, \nu ^{V})$ of Levy bridge $V^{(2)}$:
$$\begin{array}{l}
B_t^V=  b_2 \,t +c_2 \int^t_0 \beta^{v,P}_s ds + \displaystyle\int ^t_0 \displaystyle\int _{\mathbb{R}^d} \stackrel{\rightarrow}{l_2(x)} ( Y^{v,P}_s(x) -1)  K _2(dx)ds,\\

C_t^V=  c_2\,t,\\

d\nu ^V(x,t) = Y^{v,P}_t(x) K _2(dx)dt\,.
\end{array}$$
To write the characteristics for  linear combination of $X^{(1)}$ and $V^{(2)}$, we take in account the fact that the processes $X^{(1)}$ and $V^{(2)}$ remain independent under $P^v$. Then, we add the additional drift coming from the change of the measure $P^v$ into $Q^v$ and we multiply the corresponding Levy measure  by the factor $Y^v_s$.
This give us the formulas  for the characteristics.
\par The process $X$ is a $(Q^v,\mathbb{F})$-martingale if  and only if its drift term under $Q^v$ is identically equal to zero, and it gives us mentioned  identity.
$\Box$ 
\subsection{Conditional information processes}
To simplify the expression for finding of the Girsanov parameters $(\beta^{v,*} , Y^{v,*})$ of the  $f$-divergence minimal equivalent martingale measure $Q^{v,*}$, we use the notations:
 $$b=\rho _1b_1+ \rho _2 b_2,\,\,\,c= \rho _1 c_1 \,^{\top}\rho _1 + \rho_2 c_2 \,^{\top}\rho _2$$
We recall that $(b,c,K)$ are the  parameters of Levy process $X$ under "historical" measure $P$.

\thm \label{tl1}Let $u(x) = \ln (x)$ and the hypothesis $(H1)$ and $(H2)$ 
hold. If there exists a predictable process $\lambda^v=(\lambda ^v_s)_{0\leq s\leq T}$ with the values in $\mathbb{R}^d$ such that for all $s\in [0,T]$
\begin{equation}\label{fl1}
 b+c\lambda^v_s+ \rho _2 c_2\beta^{v,P}_s+\rho _2\int _{\mathbb{R}^d} \stackrel{\rightarrow}{l_2(x)} [ Y^{v,P}_s(\rho _2^{-1}x) -1] (K_2\circ\rho _2^{-1})(dx)
\end{equation} 
$$ \hspace{5cm}+ \int _{\mathbb{R}^d}\stackrel{\rightarrow}{l(x)} \frac{^{\top}\!\lambda^v_s\,\,l(x)}{1-^{\top}\!\lambda^v_s\, l(x)}K ^{v,P}_s(dx)=0,$$

and such that $1-\,^{\top}\! \lambda^v_s\,l(x) >0\,\,( K^{v,P}$-a.s.), then the Girsanov parameters of $f$-divergence minimal martingale measure $Q^{v,*}_T$ verify:
$$\beta^{v,*}_s=
 \lambda ^v_s,
\,\,\,\,\, Y^{v,*}_s(x) = \frac{1}{1-^{\top}\!\lambda^v_s\,l(x)}\,.$$
The corresponding information process $\mathcal{I}^*(v)$ is given by
\eqref{71} and the corresponding entropy is equal to \eqref{76}. If this entropy is finite, the corresponding measure will be $f$-divergence minimal equivalent martingale measure.

\pf \rm To find the Girsanov parameters of the corresponding $f$-divergence minimal martingale measure we minimise the relative entropy of $P^v_T$ given $Q^v_T$:
$${\bf I}(P^v_T\,|\, Q^v_T)= {\bf E}_{P^v_T} (\mathcal{I}_T(v))$$
with
$$\mathcal{I}_T(v)= \frac{1}{2}\int ^T_0 \,^{\top}\!\beta ^{v}_s\, c\,\beta ^{v}_s ds - \int ^T_0\int _{\mathbb{R}^d}\left( \ln Y^ {v}_s(x) - Y^{v}_s + 1\right) K ^{v,P}_s(dx) ds,$$
under constraint: for $s\in [0,T]$
\begin{equation}\label{fl2}
R(\beta ^v_s, Y^v_s)= 0\,.
\end{equation}
In this constraint, the function $R(\beta ^v_s, Y^v_s)$ is defined as follows:
$$R(\beta ^v_s, Y^v_s)= b+ \rho_2c_2 \beta^{v,P}_s + \int _{\mathbb{R}^d}\stackrel{\rightarrow}{l(x)} [Y^{v,P}_s(\rho _2^{-1}x) -1] (K_2\circ\rho _2^{-1})(dx)$$ 
$$ +c\beta ^v_s + \int _{\mathbb{R}^d}\stackrel{\rightarrow}{l(x)} (Y^{v}_s(x) -1) K ^{v,P}_s(dx)\,.$$
\par According to the traditional procedure of minimisation, we introduce the function $G$ with
$$G(\beta ^v_s, Y^v_s) = \frac{1}{2} \,^{\top}\!\beta ^{v}_s\, c\,\beta ^{v}_s   - \int _{\mathbb{R}^d}\left( \ln (Y^ {v}_s(x)) - Y^{v}_s + 1\right) K ^{v,P}_s(dx)
-^{\top}\!\lambda^v_s R(\beta ^v_s, Y^v_s),$$
where $\lambda ^v_s$ is the Lagrangian factor.
This function is convex continuously differentiable function, its extreme points are stationnary points, which  are the solutions of the equations:\\
$$\left\{\begin{array}{l}
^{\top}\!\left(\frac{\partial G}{\partial \beta_1}(\beta ^v_s, Y^v_s), \cdots \frac{\partial G}{\partial \beta_d}(\beta ^v_s, Y^v_s)\right) =c(\beta^v_s-\lambda ^v_s)=0,\\\\
\frac{\partial G}{\partial Y}(\beta ^v_s, Y^v_s)= \displaystyle\int _{\mathbb{R}^d}\left(1- \frac{1}{Y^ {v}_s(x)}- \,^{\top}\!\lambda ^v_s\,
l(x)\right)K ^{v,P}_s(dx)=0\,.
\end{array}\right.\\$$
It is clear that $\beta ^v_s=\lambda ^v_s$ is a solution of the first equation.  In general, second equation has multiple solutions, but due to the convexity of $G$, the corresponding value of the information process will be the same. One of the solutions  of the second equation is given by
$$ Y^v_s(x)= \frac{1}{1-\,^{\top}\!\lambda ^v_s\,l(x) }$$
and we assume that it is positive.
Finally, we put the expression for $\beta^v_s$ and $Y^v_s$ into the martingale condition
\eqref{fl1}, to find $\lambda ^v_s$, and, hence, $\beta^{v,*}$ and $Y^{v,*}$.
\par The convexity of the function $G$ gives
$$\hspace{2cm}G(\beta ^v_s, Y^v_s) - G(\beta ^{v,*}_s, Y^{v,*}_s)\geq $$
$$^{\top}\!\left(\frac{\partial G}{\partial \beta_1}(\beta ^{v,*}_s, Y^{v,*}_s), \cdots \frac{\partial G}{\partial \beta_d}(\beta ^{v,*}_s, Y^{v,*}_s)\right) (\beta ^v_s - \beta ^{v,*}_s) + 
\frac{\partial G}{\partial Y}(\beta ^{v,*}_s, Y^{v,*}_s)(Y^v_s -Y^{v,*}_s) =0\,.$$
To prove that the corresponding measure is $f$-divergence minimal,  i.e. $${\bf I}(P^v_T\,|\, Q^v_T)\geq {\bf I}(P^v_T\,|\, Q^{v,*}_T)\,,$$  we integrate the above inequality w.r.t. $s$ and we take expectation with respect to the measure $P^v_T$.
$\Box$

\thm \label{tl2}Let $u(x) =x \ln (x) +x -1$ and the hypothesis $(H1)$ and $(H2)$ 
are valid. If there exists predictable process $\lambda^v=(\lambda ^v_s)_{0\leq s\leq T}$ with the values in $\mathbb{R}^d$ such that for all $s\in [0,T]$
\begin{equation}\label{fl3}
 b+ c\lambda^v_s+ \rho _2 c_2\beta^{v,P}_s+\rho _2\int _{\mathbb{R}^d} \stackrel{\rightarrow}{l_2(x)} [ Y^{v,P}_s(\rho _2^{-1}x) -1] (K_2\circ\rho _2^{-1})(dx)
\end{equation}
 $$\hspace{4cm}+ \int _{\mathbb{R}^d}\stackrel{\rightarrow}{l(x)}[ \exp(\,^{\top}\! \lambda^v_s\,l(x)) -1]K ^{v,P}_s(dx)=0$$

then the Girsanov parameters of the $f$-divergence minimal martingale measure $Q^{v,*}_T$ verify:
$$\beta^{v,*}_s= \lambda^v_s,
\,\,\,\,\, Y^{v,*}_s(x) = \exp(\,^{\top}\! \lambda^v_s\, l(x))\,.$$
Moreover, the corresponding information process $I^*(v)$ is given by \eqref{83} and the Kullback-Leibler information is given by \eqref{83a}. If this  Kullback-Leibler information is finite, the corresponding measure will be $f$-divergence minimal equivalent martingale measure.

\pf \rm To find the Girsanov parameters of the $f$-divergence minimal martingale measure $Q^v_T$, we minimise relative entropy of $Q^v_T$ given $P^v_T$:
$${\bf I}(Q^v_T\,|\, P^v_T)= {\bf E}_{Q^v_T} (I_T(v))$$
with
$$I_T(v)=  \frac{1}{2}\int ^T_0 \,^{\top}\!\beta ^{v}_s\, c\,\beta ^{v}_s ds + \int ^T_0\int _{\mathbb{R}^d}\left( Y^ {v}_s(x)\,\ln (Y^ {v}_s(x)) - Y^{v}_s + 1\right) K ^{v,P}_s(dx) ds,$$
under constraint \eqref{fl2}.
 For that we introduce the function $G$ such that
$$G(\beta ^v_s, Y^v_s) =  \frac{1}{2} \,^{\top}\!\beta ^{v}_s\, c\,\beta ^{v}_s  + \int _{\mathbb{R}^d}\left(\,Y^ {v}_s(x)\, \ln (Y^ {v}_s(x)) - Y^{v}_s + 1\right) K ^{v,P}_s(dx)
-\,^{\top}\!\lambda^v_s \,R(\beta ^v_s, Y^v_s)$$
with the Lagrangian factor $\lambda ^v_s$.
This function is convex continuously differentiable function, so,  the minimum of this function is realised on the set of stationary points, which verify :\\
$$\left\{\begin{array}{l}
^{\top}\!\left(\frac{\partial G}{\partial \beta_1}(\beta ^v_s, Y^v_s), \cdots \frac{\partial G}{\partial \beta_d}(\beta ^v_s, Y^v_s)\right)=c(\beta^v_s-\lambda ^v_s)=0,\\\\
\frac{\partial G}{\partial Y}(\beta ^v_s, Y^v_s)= \displaystyle\int _{\mathbb{R}^d}\left(\ln (Y^ {v}_s(x)) -\, ^{\top}\!\lambda ^v_s
l(x)\right)K ^{v,P}_s(dx)=0\,.
\end{array}\right.\\$$
The solution of the first equation is $\beta ^v_s=\lambda ^v_s$.  We remark that second equation has multiple solutions, but the corresponding value of the information process will be the same. One of the solutions  of the second equation is given by
$$ Y^v_s(x)= \exp (\,^{\top}\! \lambda ^v_s\,l(x))\,.$$
To find $\lambda^v$, we put the expressions  for $\beta^v_s$ and $Y^v_s$ into martingale condition \eqref{fl3},  and it gives us the expressions for $\beta^{v,*}$ and $Y^{v,*}$.
\par We clearly have:
$$G(\beta ^v_s, Y^v_s) - G(\beta ^{v,*}_s, Y^{v,*}_s)\geq $$
$$^{\top}\!\left(\frac{\partial G}{\partial \beta_1}(\beta ^{v,*}_s, Y^{v,*}_s), \cdots \frac{\partial G}{\partial \beta_d}(\beta ^{v,*}_s, Y^{v,*}_s)\right) (\beta ^v_s - \beta ^{v,*}_s) + 
\frac{\partial G}{\partial Y}(\beta ^{v,*}_s, Y^{v,*}_s)(Y^v_s -Y^v_s) =0\,.$$
To show that the corresponding measure is $f$-divergence minimal, we integrate this inequality w.r.t. $s$ and we take the expectation with respect to $Q^v_T$. Then,
 $${\bf I}( Q^v_T\,|\,P^v_T)\geq {\bf I}( Q^{v,*}_T\,|\,P^v_T)\,.$$
$\Box$

\thm \label{tl3}Let $u(x) = x^p,\, p<1$ and the hypothesis $(H1)$ and $(H2)$ are satisfied. If there exists predictable process $\lambda^v=(\lambda ^v_s)_{0\leq s\leq T}$ with the values in $\mathbb{R}^d$ such that for all $s\in [0,T]$ and $q=\frac{p}{p-1}$
$$ b+ \frac{c\lambda^v_s}{q(1-q)}+ \rho _2 c_2\beta^{v,P}_s+\rho _2\int _{\mathbb{R}^d} \stackrel{\rightarrow}{l_2(x)} [ Y^{v,P}_s(\rho _2^{-1}x) -1] (K_2\circ\rho _2^{-1})(dx)$$
$$\hspace{3cm}+ \int _{\mathbb{R}^d}\stackrel{\rightarrow}{l(x)}\left[ \left(1-\frac{\,^{\top}\!\lambda^v_s\,l(x)}{q}\right)^{\frac{1}{q-1}} -1\right]K ^{v,P}_s(dx)=0,$$
and such that $1-\frac{\,^{\top}\!\lambda^v_s\,l(x) }{q}>0\,\,( K ^{v,P}$-a.s.), then the Girsanov parameters of $f$-divergence minimal martingale measure $Q^{v,*}_T$ verify:
$$\beta^{v,*}_s=
\frac{1}{q(1-q)}\lambda ^v_s,
\,\,\,\,\,\, Y^{v,*}_s(x) = \left(1-\frac{\,^{\top}\!\lambda^v_s\,l(x)}{q}\right)^{\frac{1}{q-1}}\,.$$

In addition, the Hellinger type process $h^{(q),*}(v)$ is defined by \eqref{88} and the corresponding Hellinger type integral is given by \eqref{hel2}. If this Hellinger integral is finite, the corresponding measure is $f$-divergence minimal equivalent martingale measure.

\pf \rm   To find the Girsanov parameters of the $f$-divergence minimal martingale measure $Q^v_T$, we minimise Hellinger integral of $Q^v_T$ and $P^v_T$:
$${\bf H}^{(q)}_T(v)= {\bf E}_{R^v_T} \exp (-h^{(q)}_T(v))$$
with\\\\
$h^{(q)}_T(v)= \frac{q(1-q)}{2}\int ^T_0 \,^{\top}\!\beta ^{v}_s\, c\,\beta ^{v}_s ds$
$$\hspace{2cm} - \int ^T_0\int _{\mathbb{R}^d}\left( (Y^ {v}_s(x))^q - q\,(Y^{v}_s-1) - 1\right) K ^{v,P}_s(dx) ds$$
under constraint \eqref{fl2}. For that we introduce the function $G$ via

$G(\beta ^v_s, Y^v_s) =  \displaystyle\frac{q(1-q)}{2}\,^{\top}\!\beta ^{v}_s\, c\,\beta ^{v}_s ds$
$$\hspace{2cm} - \int _{\mathbb{R}^d}\left((Y^ {v}_s(x))^q - q\,(Y^{v}_s-1) - 1\right) K ^{v,P}_s(dx)
-\,^{\top}\!\lambda^v_s R(\beta ^v_s, Y^v_s)$$
where $\lambda ^v_s$ is again the Lagrangian factor.
This function is convex continuously differentiable function, so, the  stationary points verify:\\
$$\left\{\begin{array}{l}
^{\top}\!\left(\frac{\partial G}{\partial \beta_1}(\beta ^v_s, Y^v_s), \cdots \frac{\partial G}{\partial \beta_d}(\beta ^v_s, Y^v_s)\right)= c( q(1-q)\beta^v_s-\lambda ^v_s)=0,\\\\
\frac{\partial G}{\partial Y}(\beta ^v_s, Y^v_s)=- \displaystyle\int _{\mathbb{R}^d}[ q\,(Y^ {v}_s(x))^{q-1}-q+ \,^{\top}\!\lambda ^v_s\,l(x)
]K ^{v,P}_s(dx)=0 \,.
\end{array}\right.\\$$
From the first equation we find that $\beta ^v_s=\displaystyle\frac{1}{q(1-q)}\lambda ^v_s$.  One of the solutions of the second equation is given by
$$ Y^v_s(x)= \left(1-\frac{\,^{\top}\! \lambda ^v_s\,l(x)}{q}\right)^{\frac{1}{q-1}}\,. $$
Next, we put the expression for $\beta^v_s$ and $Y^v_s$ in the martingale condition
\eqref{fl3} to find $\lambda ^v_s$ and, then, $\beta ^{v,*}_s$ and $ Y^{v,*}_s$.
\par Since $G$ is convex,
$$G(\beta ^v_s, Y^v_s) - G(\beta ^{v,*}_s, Y^{v,*}_s)\geq$$
$$^{\top}\!\left(\frac{\partial G}{\partial \beta_1}(\beta ^{v,*}_s, Y^{v,*}_s), \cdots \frac{\partial G}{\partial \beta_d}(\beta ^{v,*}_s, Y^{v,*}_s)\right) (\beta ^v_s - \beta ^{v,*}_s) + 
\frac{\partial G}{\partial Y}(\beta ^{v,*}_s, Y^{v,*}_s)(Y^v_s -Y^v_s) =0\,.$$
Then, we integrate this inequality w.r.t. $s$, we use the fact that exponential is convex function, and, finally, we take expectation with respect to $R^v_T$, in order to prove that
$${\bf H}^{(q)}_T\geq {\bf H}^{(q),*}_T,$$
i.e. that the measure $Q^{v,*}_T$ is f-divergence minimal.
$\Box$

\end{section}

%%%%%%%%%%%%%%%%%%%%%%%%%%%%%%%%%%%%%%%%%%%%%%%%%%%%%%%%%%%%%%%%%%%%%%%%%%%%%%
\begin{section}{Black-Sholes models with correlated Brownian motions}\label{s4}
%%%%%%%%%%%%%%%%%%%%%%%%%%%%%%%%%%%%%%%%%%%%%%%%%%%%%%%%%%%%%%%%%%%%%%%%%%%%%%%

\par Let $(W^{(1)}, W^{(2)})$ be independent standard Brownian motions. Let $\mu _1,\mu _2 \in \mathbb R$ and $\sigma _1>0, \sigma _2>0$. We put
$$X^{(1)}_t= \mu _1 +  \sigma _1 \,W^{(1)}_t, \,\,\,
X^{(2)}_t= \mu _2 +  \sigma _2\,W^{(2)}_t,  $$
and for the parameter $|\rho |\leq 1$, let
$$X_t= \sqrt{1-\rho ^2}X^{(1)}_t + \rho X^{(2)}_t\,.$$
Then, $X$ will be Brownian motion with drift coefficient $$\mu =\sqrt{1-\rho ^2}\mu_1+\rho \mu _2,$$  diffusion coefficient $$\sigma ^2 =(1-\rho ^2)\sigma ^2_1 + \rho ^2 \sigma ^2_2,$$ and  the correlation coefficient between $X_t$ and $X^{(2)}_t$  equal to $\rho$.
We take $ W^{(2)}_{T'}$ for conditioning instead of $X^{(2)}_{T'}$ since these two variables are in bijection.  But this replacement also implies that we should replace  $g(v)$ by $\tilde{g}(v)= \exp\{(\mu _2\,T' + \sigma _2 v\}$ in maximum utility formula.  In this setting, the law $\alpha$  is, evidently, nothing else as $\mathcal N(0,T')$.
\par We see that the hypotheses $(H1)$ and $(H2)$ are verified. In fact, the processes $X^{(1)}$ and $X^{(2)}$ are integrable, both have a strictly positive density with respect to Lebesgue measure. In particular, as well known, $W^{(2)}_t$ has a strictly positive density w.r.t. Lebesgue measure for $t>~0$:
$$f(t,x)= \frac{1}{\sqrt{2\pi t}} \exp\{-\frac{x^2}{2t}\}$$
which is $C^{1,2}_b([\epsilon , \infty[)$ for any $\epsilon >0$. Moreover, we
use normal correlation theorem to get that 
$$p^v_t=\frac{dP^v_T}{dP_T}(X) = \left(\frac{T'}{T'-\rho ^2T}\right)^{1/2}\exp \left\{-\frac{1}{2}
\left[ \frac{(v-\rho X_T)^2}{T'-\rho ^2 T} - \frac{v^2}{T'}\right]\right\}\,.$$
Then, we write  this quantity as a stochastic exponential 
\begin{equation*}
p^v_T(X)=\exp\left\{\int_0^T{\beta}_s^{v,P}dX^c_s-\frac{1}{2}\int_0^T({\beta}_s^{v,P})^2ds\right\},
\end{equation*}
where $X$ is canonical process, and we deduce that $P$-a.s. and for $t\in~[0,T]$
 \begin{equation}{\beta}_t^{v,P}=\rho \,\frac{v-\rho X_t^c}{T^{'}-\rho ^2t}\,.\label{bvp}\end{equation} 
\par After calculations, we obtain the conditional information quantities. 

%%%%%%%%%%%%%%%%%%%%%%%%%%%%%%%%%%%%%%%%%%%%%%%%%%%%%%%%%%%%%
\prop \label{t5}(cf. \cite{EV}) For entropy, Kullback-Leibler information and Hellinger type integrals 
we have:
$$\hspace{-2.5cm}{\bf I} ( P^v\,|\, Q^{v,*})= \frac{1}{2}\ln \left(\frac{T'}{T'-\rho ^2\,T}\right)+ \frac{T}{2}\left(\frac{\mu }{\sigma } + \frac{\rho v}{T'}\right)^2 - \frac{\rho ^2  T}{2 T'},$$
$${\bf I} ( Q^{v,*}\,|\, P^v)= -\frac{1}{2}\ln \left(\frac{T'}{T'-\rho ^2\,T}\right) + \frac{T T'}{2 (T'-\rho ^2T)}\left( \frac{\mu }{\sigma }+\frac{\rho v}{T'}  \right)^2 + \frac{\rho ^2 T}{2(T'-\rho ^2T)},$$
$${\bf H}_T^{(q)}(v)=\left( \frac{T'}{T' - q\rho ^2 T}\right)^{1/2}\left(\frac{T' - \rho ^2T}{T'}\right)^{q/2}\exp\left\{-\frac{q(1-q)T}{2(T'-q\rho^2 T)}\left(\frac{\mu }{\sigma } + \frac{\rho v}{T'}\right)^2\right\}\,.$$
%%%%%%%%%%%%%%%%%%%%%%%%%%%%%%%%%%%%%%%%%%%%%%%%%%%%%%%%%%%%%\\
\\
\rm Finally, to know maximum of utility, we use the Theorem 2 with $\alpha$ being $\mathcal{N}(0,T')$.  

\end{section}

%%%%%%%%%%%%%%%%%%%%%%%%%%%%%%%%%%%%%%%%%%%%%%%%%%%%%%%%%%%%%%%%%%%%%%%%%%%%%%%%%
\begin{section}{Some jump-type models}\label{s5}
%%%%%%%%%%%%%%%%%%%%%%%%%%%%%%%%%%%%%%%%%%%%%%%%%%%%%%%%%%%%%%%%%%%%%%%%%%%%%%%%%

\par Let $(W^{(1)},W^{(2)})$ be two standard Brownian motion with correlation $\rho$, $|\rho| \leq 1$. Let N be homogeneous Poisson process of intensity $\lambda >0$, independent from $(W^{(1)},W^{(2)})$. We put
$$X_t= \mu _1 t + \sigma _1W^{(1)}_t+N_t, \,\,\, t\in [0,T],$$
$$X^{(2)}_t= \mu _1 t + \sigma _1 W^{(2)}_t,  \,\,\, t\in [0,T']$$
with $T'>T$. The option will be supported by $g(X^{(2)}_{T'})$ where $g$ is measurable non-negative function on $\mathbb{R}$.
\par 
Using the same arguments as in Section \ref{s4}, we take $W^{(2)}_{T'}$ instead of $X^{(2)}$ with replacing of $g(v)$ by $\tilde{g}(v)= \exp\{(\mu _1\,T' + \sigma _1 v\}$. 
We can verify exactly in the same manner as in previous section that the hypothesis $(H1)$ and $(H2)$ are verified. Moreover, 
$$p^v_T(X) = \frac{dP^v_T}{dP_T}(X) = (\frac{T'}{T'-\rho ^2 T})^{1/2}\exp\left\{-\frac{1}{2}\left[\frac{(\sigma _1 v- \rho X^c_T)^2}{\sigma _1^2(T'-\rho ^2T)}-\frac{v^2}{T'} \right]\right\}$$ 
with $X$ canonical process corresponding to $X^{(1)}$ and $X^c$ being its continuous martingale part. Writing the last expression as stochastic exponential, we find that $P$-a.s. and  for $t\in [0,T]$
\begin{equation}\label{j5}
 \beta_t^{v,P} = \frac{\rho \,(v \sigma _1 - \rho  X^c_t) }{\sigma _1 ^2 (T'-\rho ^2 t)}\,.
\end{equation}
 We remark that $Y^{v,P}=1$ here since $N$ and $W^{(2)}$ are independent.
\par In the following lemma we give the equations for the Girsanov parameters
 $(\beta^{v,*},Y^{v,*})$ of the change of the measure $P^v$ into  $Q^{v,*}$. 
\lem \label{pj3}
The Girsanov parameters $({\beta}^{v,*},{Y}^{v,*})$ of the equivalent $f$-divergence minimal martingale measure
$Q_T^{v,*}$ are the solutions of the following equations:
\begin{enumerate}
\item for logarithmic utility and $f(x) = -\ln(x)$
$$\frac{\lambda }{\sigma _1^2} ( {Y}^{v,*}_t-1) + \frac{\mu _1}{\sigma _1^2} + 
{\beta} ^{v,P}_t - \frac{1}{{Y}^{v,*}_t}+1=0,\,\,\,\,\beta ^{v,*}_t= 1-\frac{1}{{Y}^{v,*}_t}, $$
\item for exponential utility and $f(x) = x\ln (x)-x+1$
$$ \frac{\lambda }{\sigma _1^2} ( {Y}^{v,*}_t-1) + \frac{\mu _1}{\sigma _1^2} + 
{\beta}^{v,P}_t + \ln ({Y}^{v,*}_t)=0,\,\,\,\,\,\beta ^{v,*}_t= \ln({Y}^{v,*}_t),$$
\item for power utility and $f(x) = - \frac{x^q}{q}$
$$\frac{\lambda }{\sigma _1^2}  ({Y}^{v,*}_t-1) + \frac{\mu _1}{\sigma _1^2} +  
{\beta} ^{v,P}_t + \frac{1}{1-q}[1-({Y}^{v,*}_t)^{q-1} ]=0\,\,\\,\,\,\beta ^{v,*}_t= \frac{1}{1-q}[1-({Y}^{v,*}_t)^{q-1}]\,.$$
\end{enumerate}

\pf \rm The result follows from Theorems \ref{tl1},\ref{tl2} and \ref{tl3}. For that we express $\lambda ^v_s$ in terms of $Y^{v,*}_s$, and we replace  $b$ by $\mu _1$, $c$ and $c_2$ by $\sigma^2_1$,  and we
incorporate the compensator of $N$ which is equal to $\lambda \delta _{1}$, where $\delta _{1}$ is delta-function at point 1. We take also in account that $l(1)=1$.$\Box$

We denote by  $\hat{f}$ a new convex function
related with the previous one by the relation
$\hat{f}(x) = f(x) + \frac{x^2}{2}$.
Let also $\hat{I}= (-\hat{f}')^{-1}$ be the derivative of Fenchel-Legendre conjugate $\hat{u}$ of $ \hat{f}$.

\prop \label{pj4}  Then we have the following expressions for $Y^{v,*}$ :
\begin{enumerate}
\item for logarithmic utility
$$Y^{v,*}_t= \frac{\sigma _1}{\sqrt{\lambda}}\,\hat{I}\left(\frac{\sigma _1}{\sqrt{\lambda}}\left( \beta_t^{v,P}+\frac{\mu _1}{\sigma _1^2} +1-\frac{\lambda }{\sigma _1^2}\right)\right),$$
\item for exponential utility
$$Y^{v,*}_t= \frac{\sigma _1^2}{\lambda}\,\hat{I}\left(\beta_t^{v,P}+\frac{\mu _1}{\sigma _1^2} + \ln (\frac{\sigma _1^2}{\lambda})-\frac{\lambda }{\sigma _1^2}\right) ,$$
\item for power utility
$$Y^{v,*}_t= \left(\frac{\sigma _1^2}{(1-q)\lambda}\right)^{\frac{1}{2-q}}\,\hat{I}\left(\left(\frac{\sigma _1^2}{(1-q)\lambda}\right)^{\frac{1-q}{2-q}}\left[(1-q)(\beta_t^{v,P}+\frac{\mu _1}{\sigma _1^2} -\frac{\lambda }{\sigma _1^2})+1\right]\right) \,.$$
\end{enumerate}

\pf \rm  These formulas follows directly from previous lemma. To obtain them, it is sufficient to do scaling of $Y$, i.e. introduce a new function $U$ such that $Y=cU$, then choose $c$ in a  way to express the l.h.s. of the equation via the function $\hat{I}$.
$\Box$

\prop \label{pj6} For the information quantities we have the following expressions:
$$\hspace{-2cm}{\bf I}(P^v _T\,|\, Q^v _T) = \int^T_0 {\bf E}_{P^v _T}\left[ \frac{1}{2}\sigma _1^2 (\beta^{v,*}_t)^2 -\lambda (\ln Y^{v,*}_t - Y^{v,*}_t +1)\right] dt,$$
$$\hspace{-1cm}{\bf I}(Q^v _T\,|\,P^v _T ) = \int^T_0 {\bf E}_{Q^v_T}\left[ \frac{1}{2}\sigma _1^2 (\beta^{v,*}_t)^2 +\lambda (Y^{v,*}_t\ln Y^{v,*}_t - Y^{v,*}_t +1)\right] dt,$$
$${\bf H}^{(q)}(v) = {\bf E}_{R^v _T} \exp\left\{ \int^T_0 \left(\frac{1}{2}( q(1-q) (\beta^{v,*}_t)^2 -\lambda (( Y^{v,*}_t)^q -q Y^{v,*}_t +q -1)   \right)dt \right\}\,.$$
\pf \rm  The expressions for information quantities can be obtained easily from general expressions via information processes given in Propositions \ref{p1}, \ref{p2} and \ref{p3} of Section 2 .
$\Box$\\\\
\rm  Finally, to obtain the maximum expected utility, we use, of course, the Theorem 2 with $\alpha$ being $\mathcal{N}(0,T')$.  
\end{section}

\end{document}